\documentclass[10pt,journal,compsoc]{IEEEtran}
\usepackage{verbatim}
\usepackage{color}
\usepackage{amssymb}
\usepackage{mathtools}
\usepackage{subfig}
\usepackage{algorithmic}
\usepackage{algorithm}
\ifCLASSOPTIONcompsoc
  \usepackage[nocompress]{cite}
\else
  \usepackage{cite}
\fi
\ifCLASSINFOpdf
\else
\fi
\begin{document}

%
\title{PERS: A Personalized and Explainable POI Recommender System}
%
%
%
%

\author{Ramesh~Baral and
        Tao~Li
\IEEEcompsocitemizethanks{\IEEEcompsocthanksitem Ramesh Baral is with the School of Computing and Information Sciences, Florida International University, Miami, Florida, 33199.\protect\\
E-mail: rbaral@cs.fiu.edu
\IEEEcompsocthanksitem Tao Li is with the School of Computing and Information Sciences, Florida International University, Miami, Florida, 33199.\protect\\
E-mail: taoli@cs.fiu.edu}
}

\IEEEtitleabstractindextext{%
\begin{abstract}
\label{abstract}
The Location-Based Social Networks (LBSN) (e.g., Facebook, etc.) have many factors (for instance, ratings, check-in time, location coordinates, reviews etc.) that play a crucial role for the Point-of-Interest (POI) recommendations. Unlike ratings, the reviews can help users to elaborate their opinion and share the extent of consumption experience in terms of the relevant factors of interest (aspects). Though some of the existing recommendation systems have been using the user reviews, most of them are less transparent and non-interpretable (as they conceal the reason behind recommendation). These reasons have induced considerable attention towards explainable and interpretable recommendation. To the best of our knowledge, this is the first paper to exploit the user reviews to incorporate the sentiment and opinions on different aspects for the personalized and explainable POI recommendation.  

In this paper, we propose a model termed as \textbf{PERS} (\textbf{\underline{P}}ersonalized \textbf{\underline{E}}xplainable POI \textbf{\underline{R}}ecommender \textbf{\underline{S}}ystem) which models the review-aspect category correlation by exploiting deep neural network, formulates the user-aspect category bipartite relation as a bipartite graph, and models the explainable recommendation using bipartite core-based and ranking-based methods. 
The major contributions of this paper are:
(i) it models users and locations based on the aspects posted by user via reviews,
(ii) it exploits a deep neural network to model the review-aspect category correlation, 
(iii) it provisions the incorporation of multiple contexts (e.g., categorical, spatial, etc.) in the POI recommendation model, 
(iv) it formulates the preference of users' on aspect category as a bipartite relation, represents it as a location-aspect category bipartite graph, and models the explainable recommendation with the notion of ordered dense subgraph extraction using bipartite core-based and ranking-based approaches,  
and (v) it evaluates the generated recommendation with three real-world datasets.
\end{abstract}

}

\maketitle

\IEEEdisplaynontitleabstractindextext

%
\IEEEpeerreviewmaketitle


\IEEEraisesectionheading{\section{Introduction}\label{sec:introduction}}
Most of the existing e-commerce systems (e.g., Amazon~\footnote{\tiny{www.amazon.com}}, etc.) have been facilitating users to share their consumption experience via ratings and review texts. 
The LBSNs (e.g., Facebook~\footnote{\tiny{www.facebook.com}}, Twitter~\footnote{\tiny{www.twitter.com}} etc.) have also been a useful platform to share consumption experience on different factors of interest (for instance, price, service, accessibility, product quality, and so forth). For instance, the review text \textit{"The food was really awesome but the front-desk service was really bad"} implies a positive experience of the reviewer towards the food quality and opposite for the customer service. Such experiences from a real customer have been crucial in the purchase decision for potential customers, and product improvement for manufacturers. 

Despite the usefulness, the reading time, and uniform interpretability of the review texts have been a major concern.
It would have been easier if one can summarize and explain the opinions on key aspects, for instance, (i) \textit{place A} has a \textit{good} rating for \textit{food}, (ii) \textit{place B} is \textit{renowned} for \textit{cleanliness}, and so forth. Though a dedicated community has been focusing on the extraction of such aspects, opinions, summary~\cite{mason2016microsummarization}, 
it has been less explored in the recommendation domain. The aspect-based summarization can also be used for explanation of recommendation. 

The research on the exploitation of different factors of LBSN for an efficient recommendation has been quite popular in the last decade~\cite{yuan2013time, zhang2016collaborative}. Most of the studies have just focused on non-text attributes, such as the categorical, temporal, spatial, and social aspects~\cite{baral2016maps, baral2016geotecs, baral2017exploiting}.
Most of the existing systems have also been concealing the reasons behind the recommendation and have been less transparent and less interpretable (i.e. the factors used to get the recommendation are hidden from end users). Contrary to that, some of the studies~\cite{vig2009tagsplanations, symeonidis2008providing, tintarev2007survey} have already claimed the usefulness of explainability on the persuasiveness of users towards real-world systems.
With the abundance of LBSN datasets and affordable computing cost, the research community has anticipated the necessity of incorporation of additional attributes for a generic, scalable, and explainable recommendation. 
Some of the motivating studies~\cite{tintarev2012evaluating, gedikli2014should, vig2009tagsplanations} have convincingly emphasized the importance of explanations.
The similarity-based approaches ~\cite{herlocker2000explaining, bilgic2005explaining} have proposed the user-based neighbor style (e.g., \textit{users with similar interest have purchased the following items}) explanations. The item-based neighbor style (e.g., \textit{items similar to you viewed or purchased in the past}), influence style (how the users' input have influenced the generation of recommendation), and keyword-style (items that have similar content to purchase history) can be some other variants of explanations. 

To the best of our knowledge, this paper is the first to explore the problem of aspect-based personalized explainable POI recommendation. 
There are many factors that make this problem challenging and interesting. First of all, the aspect extraction from ambiguous and noisy text itself is a difficult task. As there can be many aspect terms, efficiently organizing them into relevant categories (e.g., food, service, etc.) is also nontrivial.
The task of personalized recommendation is challenging as we need to model individual user preferences. The aspect-based personalized explanation generation becomes challenging as it needs to deal with the sentiments associated with the aspects, and also the individual user preferences and item features to get the relevant explanation. 

The ease of adaptation of arbitrary continuous and categorical attributes in a scalable manner makes the Convolutional Neural Networks (CNN) a good candidate for classification problems (e.g., ~\cite{kim2014convolutional, collobert2011natural}). This also makes them ideal for review-aspect category correlation problem.
This paper first formulates the problem of review and aspect category correlation using Convolutional Neural Networks. This simplifies the process of mapping the user sentiments to the (location, aspect category) tuples and modeling the users' aspect category preferences as the aspect category-location bipartite relation. We represent such a bipartite relation using a bipartite graph and extract the ordered aspect category preference of a user using bipartite core extraction, ranking-based, and dense subgraph extraction-based methods to generate an explainable POI recommendation. 
The core contributions of this paper are:
\begin{enumerate}
	\item it exploits the aspects from review text and categorizes them into different aspect categories,
	
	\item it formulates the review-aspect category correlation using deep neural network (see Table~\ref{tab:aspect_categories} for detail) and models users and places using those aspect categories and other different attributes (for instance, categorical, spatial, social, etc.) for efficient POI recommendation, 
	
	\item it formulates the user preferences as an ordered aspect category-location bipartite relation, represents it as a bipartite graph, and exploits the bipartite core, ranking, and shingles-based dense subgraph extraction methods to generate explanations for the personalized POI recommendation, and
	
	\item it evaluates the generated recommendation with three real-world datasets.
	
\end{enumerate}
As an important by-product, our model can implicitly identify the user communities and categorize them by their preferred sets of aspects.

\section{Related Research}
\label{sec:rel_res}
The problem of aspect extraction from review text has been there for a while~\cite{li2012opinion, zhai2011clustering} and has been used in various interesting problems (e.g., rating prediction~\cite{mcauley2013hidden}, 
aspect-sentiment summarization~\cite{titov2008joint, moghaddam2011ilda, jo2011aspect}, recommendation~\cite{zhang2015orec}, etc.). To the best of our knowledge, the exploitation of aspects for explainable POI recommendation has barely been explored. 

\subsection{Aspect-based approaches}
Yang et al.~\cite{yang2013sentiment} exploited the tips from Foursquare to extract user preferences. They relied on a sentiment lexicon (e.g., SentiWordNet)-based approach and defined the preferences based on the tips, check-ins, and social relation to generate recommendations.
They did not fully exploit the preferences at aspect level and also had no provision of recommendation explanation.
Wang et al.~\cite{wang2015semantic} exploited multi-modal (i.e. text, image, etc.) location semantic similarity. The topics extracted using LDA~\cite{blei2003latent} was used to find similar locations. 
Their model also did not focus on the aspect level preference modeling, and recommendation explanation.

Zhang et al.~\cite{zhang2015orec} exploited user opinions from the tips and fused tip polarities, social links, and geographical information for POI recommendation. 
Though their fused model was claimed efficient for polarity prediction of tips, and for location recommendation, the recommendation was not generated for individual aspects, and had no provision for the explanation.
Covington et al.~\cite{covington2016deepyoutube} used DNN for Youtube video recommendation. 
They first applied a module to filter out potential candidates and then used a deep network for the recommendation.
They incorporated different factors, such as users' activity history, demographics, etc., but did not incorporate the opinions from user comments, and also did not have any provision of recommendation for each aspect category. 
Manotumruksa et al. \cite{manotumruksa2016modelling} used word embedding to model the users and venues. 
Though they used context attributes from review texts, provision for additional attributes (such as location category, check-in time, etc.), and sentiment polarity remained unexplored. A recent study from Bauman et al.~\cite{bauman2017aspect} exploited user reviews to enhance the recommendation by incorporating the aspects.

Recently, Zheng et al.~\cite{zheng2017joint} adapted~\cite{collobert2011natural} to exploit the user reviews and to
map the user and item feature vectors into same space and estimated the user-item rating.
Our model has following major differences from~\cite{zheng2017joint}: 
(i) it uses the sentiment polarity of reviews at the sentence level rather than the whole review text,
(ii) it learns to classify each review sentence into aspect categories, and models users and places using these aspect categories and embedding of additional features (e.g., the location category, check-in time, etc.), and 
(iii) it efficiently exploits bipartite core extraction and ranking method to segregate the aspect categories and relevant locations for explainable recommendation. 
\subsection{Explanation-based approaches}
Chen et al.~\cite{chen2016learning} 
personalized ranking based tensor factorization model and used phrase-level sentiment analysis across multiple categories.
They 
extracted aspect-sentiment pairs from review text, and used Bayesian Personalized Ranking~\cite{rendle2009bpr} to rank the features mentioned by a user in her review text. 
Finally, the feature wise preference of a user was derived using the user-item-feature cube and rank of the feature.
A similar study~\cite{zhang2014explicit} used matrix factorization to estimate the missing values and the recommendation was made by matching the most favorite features of a user with the matching properties of the items. They used simple text templates to generate a feature-based explanation of positive and negative recommendations for an item. However, incorporation of additional features (for instance, location category) was not explored in their research.

Lawlor et al.~\cite{lawlor2015opinionated} exploited sentiment wise explanation to explain why a place might or might not be interesting to a user. For every aspect, they also compared the recommended place to the alternatives and provided the explanation (for instance, better (worse) than 90\% (20\%) of alternatives for \textit{room quality} (\textit{price}), etc.). However, they simply relied on the frequency of aspects of locations and the users to get such relation and the incorporation of additional features for location prediction remained unexplored. He et al.~\cite{he2015trirank} modeled the user-item-aspect as a tri-partite graph and used the graph-based ranking algorithm to find the most relevant aspects of a user that match with the relevant aspects of places. The common relevant aspects were used in the explanation.

We have found that only some of the models have fused few additional attributes (e.g., social), whereas most of them had no provision for additional attributes. Most of the studies were tightly coupled to the aspects and their sentiments. They analyzed the influence of all the aspects together. The influence of aspects among each other can have some adverse impact on the personalized recommendation. For instance, a place that is good in "Price" category might be opposite in "Service" category. A user who just cares about "Price" aspect might ignore some "Service" related problems in that place. So we need to minimize the influence of aspects among each other.
This is really crucial for aspect-based recommendation systems and to the best of our knowledge, none of the existing studies have explored this direction. We attempt to fill this gap by the concept of bipartite graph and bipartite core (dense subgraph) extraction. For a user, the most dense subgraph represents the most preferred aspect and the places that are most popular for this aspect. The dense sub graph extraction is followed by disconnecting the edges within the dense subgraph. By doing this we can have less interference from the aspects which were already discovered in previous dense subgraphs. This claim is also supported by our evaluation where one of our model \textbf{PERS-Core} performs slightly better than our another model \textbf{PERS-Rank} (see Section~\ref{sec:methodology_bipartitecores} for detail).

\section{Methodology}
\label{sec:methodology}
The overview of proposed system is illustrated in~Figure~\ref{fig:arch_combined}.

\subsection{System architecture}
\label{ssec:methodology_arch}

\begin{figure*}[!t]
	\centering
	\subfloat[Overview of review classification module]{\includegraphics[width=0.9\columnwidth,height=0.6\columnwidth]{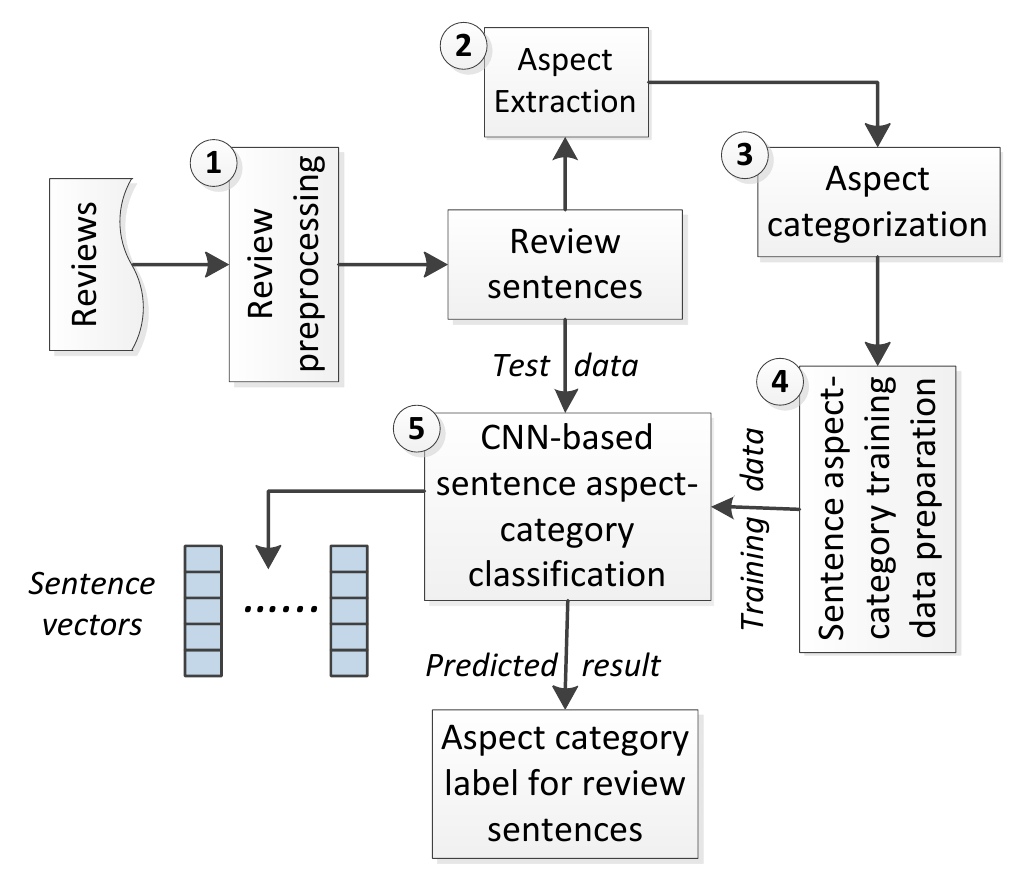}
		\label{fig:arch_a}}
	\hfil
	\subfloat[Overview of recommendation module]{\includegraphics[width=0.9\columnwidth,height=0.6\columnwidth]{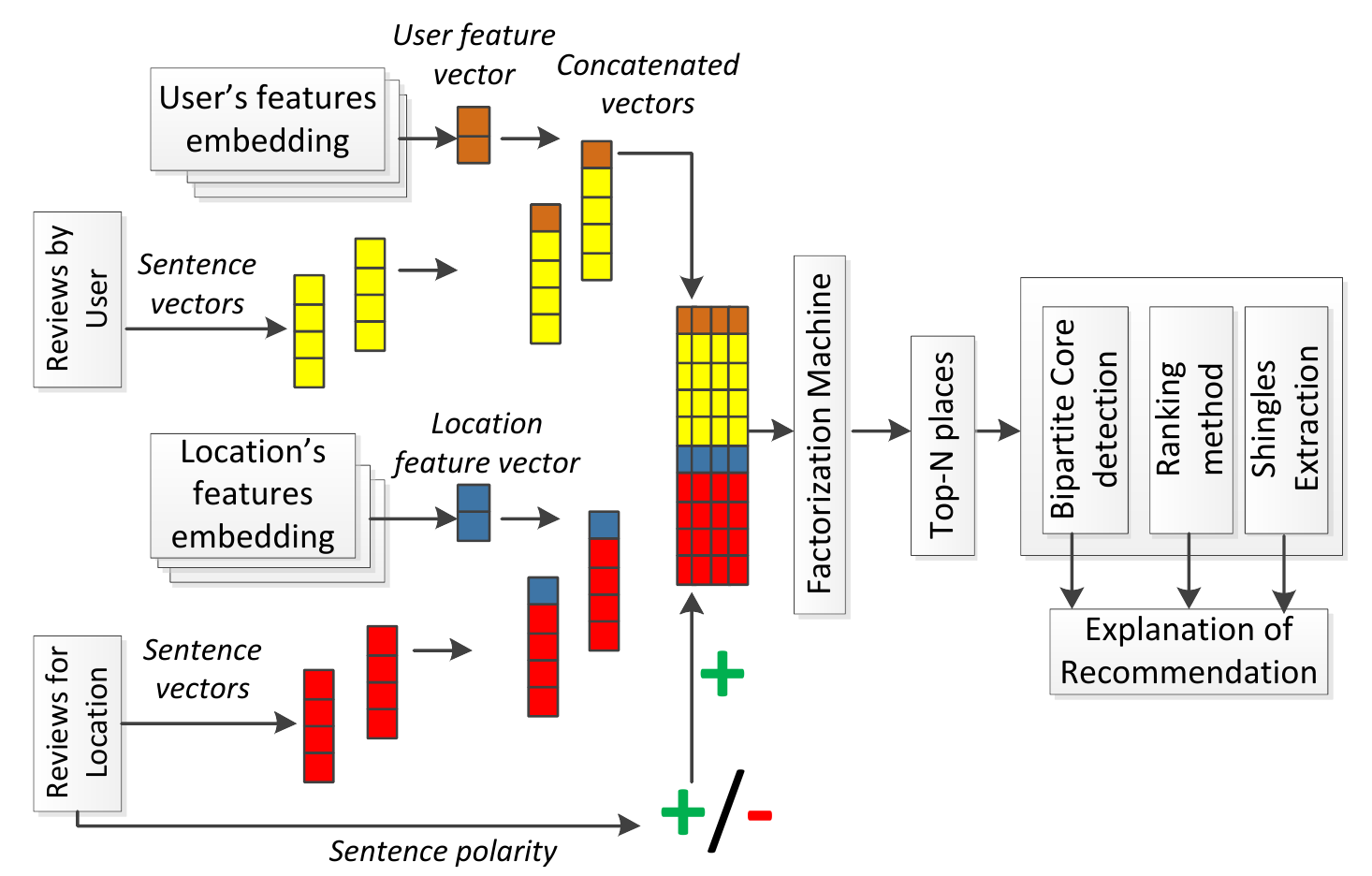}
		\label{fig:arch_b}}
	\caption{High level overview of system architecture}
	\label{fig:arch_combined}
\end{figure*}

The core features of the proposed system are:
\begin{enumerate}

	\item Review preprocessing: 
	For every user, the review texts posted for every location are segregated by the ratings and are organized in separate text files. As the user rating is for whole review text rather than a single sentence, we use VADER~\cite{hutto2014vader} to correct the sentiment polarity of individual review sentences. 
	This approach is reasonable because every review sentence might deal with different aspects and might have different sentiment.
	
	\item Aspect extraction:
	The pre-processed review sentences are fed to the aspect extraction module to extract the aspects. 
	We exploit a simple two-step process for the aspect extraction.
	The first task filters out the nouns and noun phrases using some experimentally set frequency threshold. 
	It has been found that most of the reviews focus on a set of topics and this approach can easily capture such topics~\cite{moghaddam2010opinion}.
	The second task is to use a rule-based approach~\cite{zhang2014aspect} that adapts the dependency parsing~\cite{manning2014stanford} to handle the aspects missed in the previous step. 
	
	\item Aspect-categorization:
	As there can be numerous aspects, for the sake of ease of computation, we narrow down them to few well-known aspect categories. Some of the popular aspect categories (see Table~\ref{tab:aspect_categories}) are considered. The aspects and their synsets from WordNet~\cite{fellbaum1998wordnet} are used to assign the best matching aspect categories. We select top 3 synsets to handle ambiguity of aspect words and to capture the relevant aspect categories.
	
	\item Sentence-aspect-category training data preparation:
	The review text (after aspect extraction) is labeled by the aspect category which has the closest match to its aspects. 
	The distance between aspect words and these aspect category words (and their synonyms) from the WordNet~\cite{fellbaum1998wordnet} are used to assign the closest possible label. As we assign top 3 matching synsets, a single aspect term can have three matching aspect categories. The default category "NONE" is applied for aspects not matching to either of the aspect categories. The sentences with multiple aspect terms get multiple aspect categories.	The performance of the training data preparation module is defined in the evaluation section (see Section ~\ref{sec:evaluation}).
	
	\item CNN-based sentence-aspect-category classifier: 
	The review-aspect category correlation module in our proposed system is a simple binary classifier (see Figure~\ref{fig:arch_a}) that 
	learns to classify a review sentence into different aspect categories. 
	As we have more than two aspect categories, we used a simple CNN based sentence classifier (see Section~\ref{ssec:methodology_deep_network}) for each of the aspect categories. This binary classifier can classify a review sentence to an aspect category or not to that aspect category.
	The input to this classifier is word embedding~\cite{mikolov2013distributedword2vec} (which maps every word to a uniform size vector in a latent feature space) of review sentences. 
	
	For every user, we get a set of sentence feature vectors which is simply an embedding of her preferred aspects. 
	Similarly, for every location, the sentence feature vector is simply an embedding of the aspects which were specified in its reviews.
	As a place might be positively or negatively mentioned for some aspects, we also use the sentence polarity to identify the user's sentiment polarity for the location's feature vector.
	The CNN-based classifier takes the user review sentences and identifies their aspect categories. The same process is repeated for the reviews made for every location. This can be easily extended with a multi-label CNN classifier that can operate on multiple aspect categories at a time. The outcome of this is a bipartite relation between user review and the aspect categories. This bipartite relation can be used to define the location-aspect category tuples and user-aspect category tuples. Such a bipartite relation can be exploited to model user preferences via ordered aspect category-place relation using bipartite graph and extraction of dense subgraphs from such graph (see Section~\ref{sec:methodology_bipartitecores} for details).
	
	
	
	\item Recommendation generation: 
	This POI recommendation model is termed as \underline{\textbf{D}}eep \underline{\textbf{A}}spect-based \underline{\textbf{P}}OI recommender (\textbf{DAP}). Besides the review text, we also incorporate additional features (for instance, categorical, spatial, etc.). For every user, such vectors are concatenated to the review vector of the place (the review made by this user) obtained from the CNN-based classifier. The same process is repeated for every location. 
	Every user tends to mention some opinion on preferred aspects in her reviews and every place is mentioned about the aspects it was reviewed for. 
	A positive review rating implies positive experience on relevant aspects and a negative review rating implies the opposite.
	
	We formulate the recommendation problem as a matrix, whose rows represent a user, location, location category, and elements of other feature vectors. 
	For each row, the check-in flag of a user to a location is treated as the target.
	For instance, if a user $u_i$ has review vector as $<ue_1,ue_2,....,ue_m>$, a place $l_j$ has its vector as $<le_1,le_2,...,le_n>$, and the place $l_j$ has positive ratings for the preferred aspects of user $u_i$, then a row in the design matrix is obtained simply by concatenating the user vector, location vector, and feature vectors, and is defined as:  $\overrightarrow{u_i,l_j} = <ue_1, ue_2, ...., ue_m$ $,le_1, le_2,..., le_n,$ $fe_1, fe_2, .... fe_k, 1>$, where $ue_a$, $le_a$, and $fe_a$ are the $a^{th}$ item (a real-valued number) of the vector of the user, place, and other features. 
	The last element \textit{1} represents the check-in flag for the user-place tuple in the training data and represents the score to be estimated for the test data. For the places that have negative rating for the preferred aspects of a user, the check-in flag is set to \textit{0}. For user $u$, we use the feature vector as $<r_{cat_1}, r_{cat_2},...,r_{cat_k}, r_{soc},r_{dist}>$, where, $r_{cat_1} = \frac{\sum\limits_{\mathclap{l.cat=cat_1}}V_{u}(l)}{\sum\limits_{\mathclap{l' \in u_L}}V_{u}(l)}$ is the ratio of total check-ins made to places with category $cat_1$ to that of all check-ins, $r_{soc} = \frac{\sum\limits_{\mathclap{l \in u_{f_L}}}V_{u}(l)}{\sum\limits_{\mathclap{l' \in u_L}}V_{u}(l')}$ is the ratio of total check-ins made on places visited due to social influence to that of all check-ins, and $r_{dist} = \frac{\sum\limits_{\mathclap{dist(l)\le \epsilon}} V_{u}(l)}{\sum\limits_{\mathclap{l'\in u_L}}V_{u}(l')}$ is ratio of total check-ins on places within a threshold distance $\epsilon$ (from users home or work place) to that of all check-ins.
	
	A factorization machine~\cite{rendle2012factorization} is exploited to estimate the value of the check-in flag for every user-place tuple. As the factorization machine has the ability to deal with additional features, a user-place pair can have multiple rows but just one row for one user-place-features tuple. So, the prediction is already somewhat personalized for the user-place-feature tuple. The top-N scorers from factorization machine are further filtered out based on the preferred aspects of user and the polarity of those aspects for places. The remaining top scorers are the recommendations made to the user.
	The high-level overview of the recommendation module is illustrated in Figure~\ref{fig:arch_b}.
	
	\item Explanation of recommendation: 
	After getting the place-aspect category bipartite relation from CNN-based classifier, we can represent the user-aspect category preference as a bipartite graph and can generate the recommendation explanation by extracting the most dense subgraphs from this bipartite graph. We propose three different methods- the first one is bipartite core-based, the second one is ranking-based, and the third one is shingles extraction-based methods for explanation generation. 
	In bipartite core extraction, the HITS algorithm~\cite{kleinberg1999authoritative} is applied to extract the most dense sub graph (the primary bipartite core) and in ranking-based method the Page Rank algorithm~\cite{page1999pagerank} is used (see Sec.~\ref{sec:methodology_bipartitecores} for detail). 
	
\end{enumerate}
\begin{table}[h!]
	\begin{tabular}{|c|c|}
		\hline 
		\rule[-1ex]{0pt}{2.5ex} \textbf{Category} & \textbf{Example}\\ 
		
		\hline 
		
		\rule[-1ex]{0pt}{2.5ex}  Price & cheap, deals, coupons, cost\\ 
		
		\hline 
		
		\rule[-1ex]{0pt}{2.5ex}  Food & food quality, food variety, free breakfast\\ 
		
		\hline 
		
		
		
		\rule[-1ex]{0pt}{2.5ex}  Service & serving time, friendly staffs\\ 
		
		\hline 
		
		
		
		\rule[-1ex]{0pt}{2.5ex}  Amenities & comfort, laundry, security, free parking, free WiFi\\ 
		
		\hline 
		
		\rule[-1ex]{0pt}{2.5ex}  Accessibility & near, disability access, information on web\\ 
		
		\hline 
		
		\rule[-1ex]{0pt}{2.5ex}  Others & security, pet friendly\\ 
		
		\hline
		
	\end{tabular} 
	\caption{Aspect categories}
	\label{tab:aspect_categories}
\end{table}

\subsection{Deep Network}
\label{ssec:methodology_deep_network}
The formulation of review-aspect\_category correlation module is inspired from the sentence classifier~\cite{kim2014convolutional}.
Our model constitutes of a CNN, an activation function, and max-pooling layer to generate the feature vectors for users and items. 
The feature vector of a user represents the preferred aspects. Similarly, the feature vector of a location represents aspects which were mentioned in its reviews. 
The sentiment of reviews is not used during the classification but is used during the recommendation.

This binary classifier learns to classify a review sentence into an aspect category. 
For instance, it is trained to predict if a sentence is related to "\textit{Price}" aspect category or not. 
This process is repeated for every aspect category and a default category "NONE" is assigned if the model fails to classify a sentence into either of the aspect categories. 
We consider one sentence at a time because every review sentence might be related to different aspect category.
This trained model is utilized to predict the aspect categories of review sentence of user and places.


\begin{figure}[!h]
	\centering
	\includegraphics[width=0.8\columnwidth,height=0.6\columnwidth]{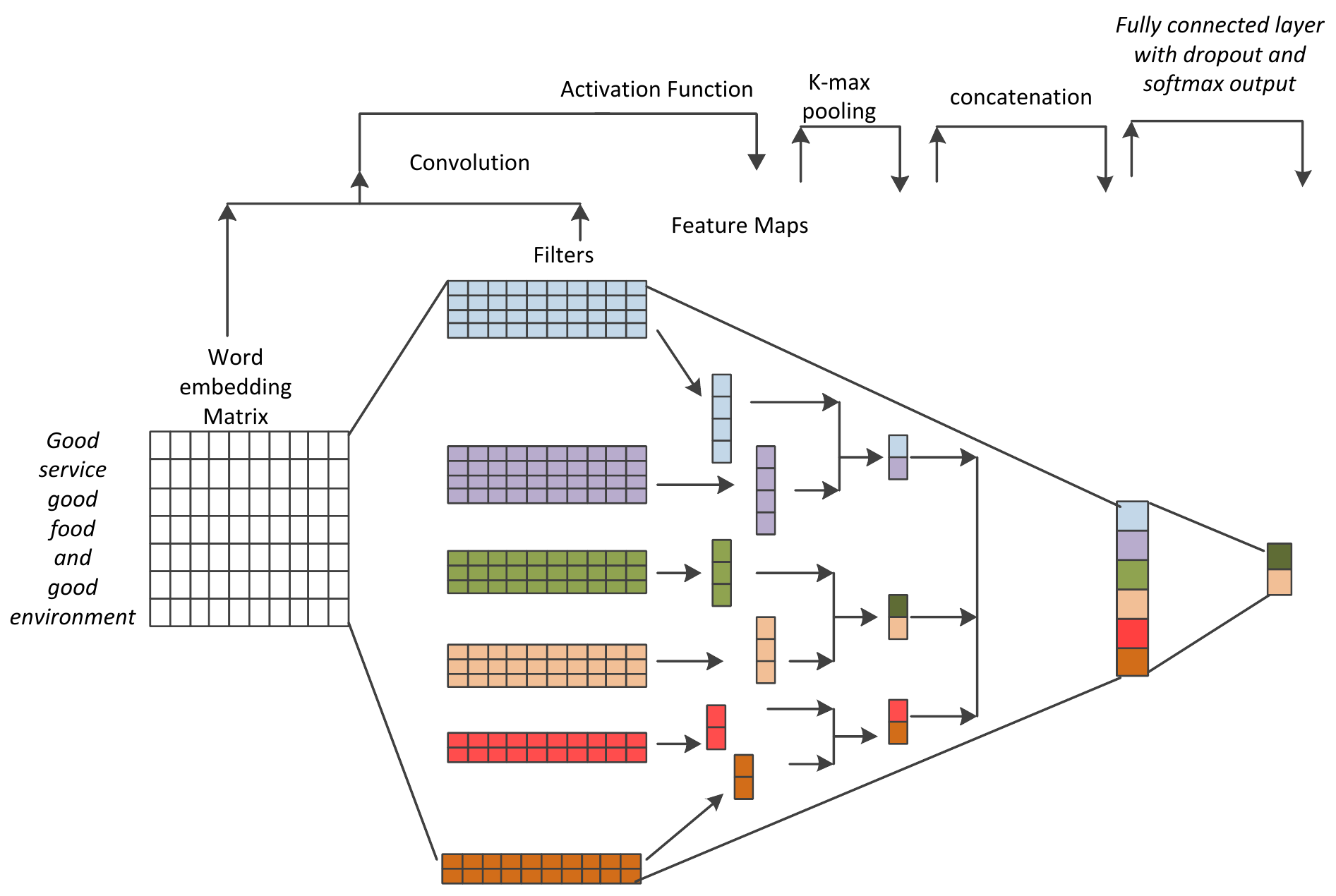}
	\caption[Deep Network]{Proposed Deep Network}
	\label{fig:deep_network}
\end{figure}

The high-level overview of the neural network-based classifier is illustrated in Figure~\ref{fig:deep_network}. The core features of the network are explained below:
\begin{itemize}
	\item Convolution Layer: The word embeddings~\cite{mikolov2013distributedword2vec} of the review sentences are input to the network.
	The convolution layer processes the embeddings of the review sentences.
	The rows are padded with zeros to maintain uniform size and a wide convolution is applied. 
	Within the convolution, an activation function (based on rectified linear units -ReLU) is applied. 
	In general, the input matrix is represented as a set of regions and the $k^{th}$ filter ($w^{(1,k)}$) of the first layer generates a feature map using the following relation:
	\begin{equation}
	z_{i,j}^{1,k} = \chi(\sum\limits_{s=0}^{r_{k}-1} \sum\limits_{t=0}^{r_{k}-1} w_{s,t}^{(1,k)}.z_{i+s,j+t}^{(0)}+b^{(1,k)}),
	\end{equation}
	where $r_k$ is the size of the $k^{th}$ filter, $\chi$ is the activation function using a ReLU, $z_{i,j}$ is the $j^{th}$ feature map in the $i^{th}$ convolution layer, and $b^{(i,j)}$ is the $j^{th}$ bias term of the $i^{th}$ convolution layer.
	
	\item Max-pooling Layer: Not all the rows of feature maps might be interesting. This is the reason behind max-pooling which selects only the k-max rows from each of the feature maps. These k-max rows are concatenated together to get a single vector. The max-pooling functionality is represented as:
	\begin{equation}
	z_{i,j}^{(2,k)} = \max\limits_{0 \le s \le d_k} \max\limits_{0 \le t \le d_{k'}} z_{i.d_k + s,\space j.d_{k'}+t},
	\end{equation}
	where $d_k$ and $d_{k'}$ are the width and length of the $k^{th}$ filter. 
	To make it simple, we use the filters of same width as the input matrix. The stride of size 1 is used.

	\item Fully connected Layer: 
	The feature vectors generated from the previous step are then fed to a fully connected layer. In this phase, the dropout regularization is applied and the input sentence is classified using a softmax function.
	
\end{itemize}

\subsection{Factorization Machine}
\label{ssec:methodology_factorization_machine}
The Factorization Machine~\cite{rendle2012factorization} formulates the prediction problem as a design matrix $X \in \mathbb{R}^{n\times p}$. 
The $i^{th}$ row $\vec{x}_{i} \in \mathbb{R}^{p}$ of the design matrix defines a case with $p$ real-valued variables. The main goal is to predict the target variable $\hat{y}(\vec{x})$ using~Equation~\ref{eq:fact_1}. 
The proposed recommendation module is formulated as a sparse matrix. The rows of the matrix are generated by concatenating the embeddings of a user, location, location category, and potentially other attributes relevant to the user-location tuple. 
We consider the check-in flag as the target variable for each row. 
The proposed model is operated with the following objective function:
\begin{equation}
\label{eq:fact_1}
\hat{y}(\vec{x}) = w_{0} + \sum\limits_{i =1}^{n} w_{i}x_{i} + \sum\limits_{i=1}^{n}\sum\limits_{j=i+1}^{n} <\vec{v}_i,\vec{v}_j> x_{i}x_{j},
\end{equation}
where $w_0$ is the global bias, $\vec{x}$ is a concatenated vector, $n$ is the size of the input variables,
$
<\vec{v}_i,\vec{v}_j> = \sum\limits_{f=1}^{k}v_{i,f}.v_{j,f},
$
and $k$ is the dimensionality of factorization. 
The Factorization Machine can learn latent factors for all the variables, and can also allow the interactions between all pairs of variables. 
This makes them preferably a good candidate to model complex relationships in the data.

\section{Explanation of recommendation}
\label{sec:methodology_bipartitecores}
The location-aspect\_category bipartite relation obtained from Section~\ref{ssec:methodology_arch} can be represented as a bipartite graph and the ordered preference of user on aspect categories can be modeled by extracting the dense subgraphs from the bipartite graph. The aspect category of the dense subgraph can be used for explanation of recommendation.
We propose bipartite core-based, and Page Rank-based methods to extract the dense subgraphs.
\subsection{Bipartite Core Extraction}
\label{ssec:methodology_bipartitecores:num1}
A k-core of a graph is a maximal connected subgraph in which every vertex is connected to at least k other vertices. The k-core analysis is commonly used as part of community detection algorithms, dense subgraph extraction, and in dynamic graphs.
Our method for bipartite core detection formulates the problem as a graph where every node is assigned two scores - hub score, and an authority score~\cite{kleinberg1999authoritative}. 
The hub score and authority score of a node are defined in terms of the outgoing and incoming edges respectively. 
The hub score ($h_i$) of a node is proportional to the sum of authority scores of the nodes it links to. The authority score ($a_i$) of a node is proportional to the sum of hub scores of the nodes it is linked from.
Given the initial authority and hub scores of all the nodes, the scores are iteratively updated until the graph gets converged.

\begin{figure}[h!]
	\centering
	\includegraphics[width=0.7\linewidth, height=0.30\linewidth]{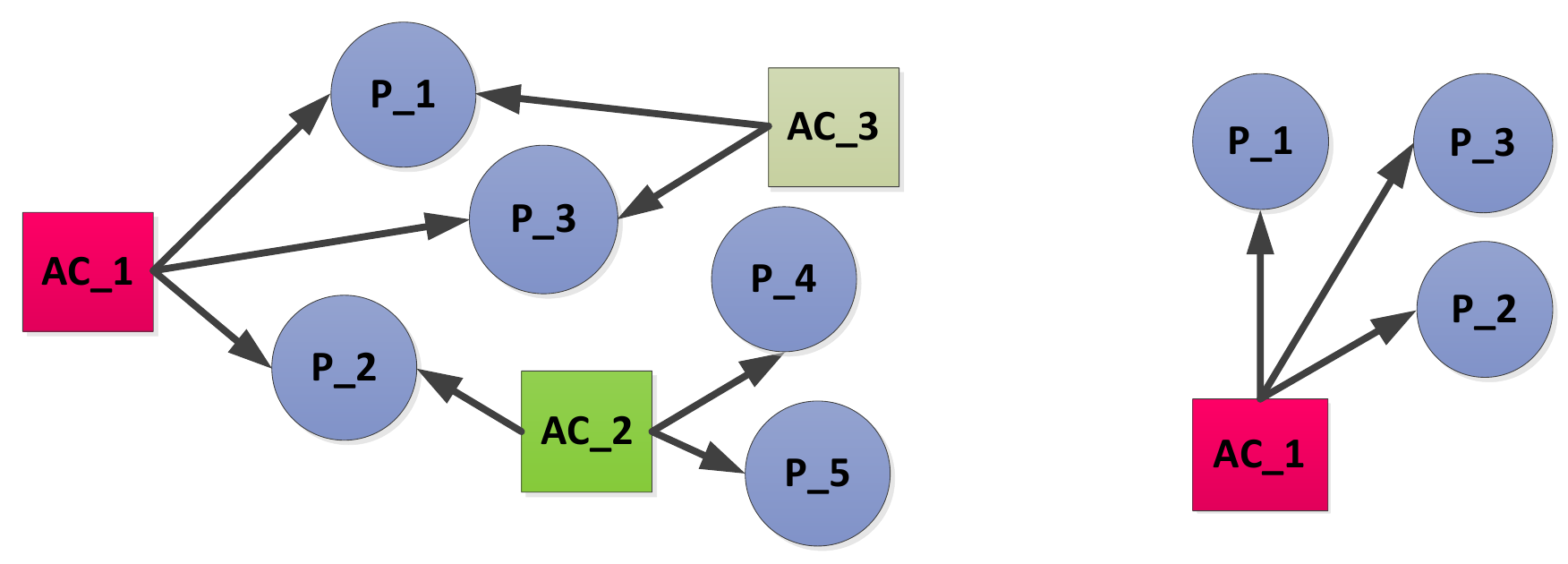}
	\caption[Place\_Aspect Category Graph]{Place Aspect\_Category Graph ($AC_k$ = aspect categories, $P_i$ = places) (Left graph is the bipartite graph and the right subgraph denotes the primary bipartite core)}
	\label{fig:hub_authority}
\end{figure}

For a given user, we consider all the recommended places as the seed nodes and connect them to the aspect category nodes for which they were positively reviewed. This gives us a graph as shown in Fig~\ref{fig:hub_authority} (left graph).
We calculate the eigenvectors of the adjacency matrix of the graph to identify the primary eigenpair (largest eigenvalue). The eigenvalue is used as a measure of the density of links in the graph. The iterative algorithm gives the largest eigenvalue (primary eigenpair). The primary eigenpair corresponds to the primary bipartite core (which represents the most prevalent set of location-aspect\_categories) and non-primary eigenpairs correspond to the secondary bipartite cores (which represent less prevalent set of location-aspect\_categories). 
After finding the primary core, the edges relevant to this core are removed and the process is repeated on residual graph to get the next prevalent bipartite cores. 
The aspect category is used in the explanation of recommendation related to those places.  
As shown in Figure~\ref{fig:hub_authority}, the left graph is a sample bipartite graph with some aspect categories and the relevant places. 
The most dense subgraph (right subgraph in Figure~\ref{fig:hub_authority}) with nodes AC$_1$, P$_1$, P$_2$, and P$_3$ are extracted as the primary bipartite core.

\subsubsection{Explanation generation from bipartite cores:}
\label{sssec:methodology_bipartitecores:num2}
Basically, a bipartite core consists of densely connected nodes and resembles the set of nodes that represent the set of place nodes which are mostly known for the relevant aspect category nodes. This model is termed as \textbf{PERS-Core} in which we use the aspect category from the bipartite cores and associate it with the explanation of recommendation related to the place nodes in the bipartite cores. 

For instance, the review text "\textit{Impressive free hot \textbf{breakfast} and free \textbf{parking}}", implies that the user cares about the \textbf{"food"} and \textbf{"amenities"} aspects. 
The bipartite cores for this user are based on her preferred aspects as she has mentioned those aspects frequently in her reviews. 
As an example, given the place "Hyatt Regency", and "The Setai Miami Beach" have good ratings for the "Food" aspects, they are included in the primary bipartite core (i.e. related to "Food") and the explanation can be provided as:
\newline
\textbf{Recommended Place}: Hyatt Regency, The Setai Miami Beach, ...
\newline
\textbf{Explanation}: Popular for Food.

The primary bipartite core includes the places and aspect categories that are most densely connected. If the places were reviewed for multiple aspect categories, they should have links to multiple aspect categories as well. The removal of edges within the primary bipartite core will still leave the nodes connected to other aspect category nodes which belong to the secondary bipartite cores. The bipartite cores are presented in the order (primary, secondary, etc.).

\subsection{Ranking Method}
\label{ssec:methodology_bipartitecores:num3}
This model is termed as \textbf{PERS-Rank} and uses the frequency of usage of an aspect category to a place. The places recommended to a user are taken and a graph with places and their relevant aspect category nodes is created. The weight of a place-aspect\_category edge indicates the number of times the place is mentioned for the aspect\_category. 
As most of the places with same category can have common aspects, we incorporate the categorical context in our ranking model to facilitate the categorical context propagation across place nodes. 
For simplicity, the edge between place nodes with same category is unit weighted. The categorical ranking relation is then defined as:

\begin{equation}
\pi(i) = \frac{1-d}{N} + d*\sum\limits_{(j,i) \in E} \frac{\pi(j)*W_{j,i}}{O_j},
\end{equation}
where $\pi(i)$ is the rank of a node \textit{i}, d (=0.85) is damping factor, N is number of nodes in the graph, E is set of edges in the graph, $W_{j,i}$ is weight of the edge (j, i), and $O_j$ is number of outgoing links from node \textit{j}.
The ranks are iteratively updated till the graph is converged. 
The highest ranking aspect category node and its neighbors give the aspect categories and the places that were noted for this aspect category. Similarly, the other higher ranking aspect category nodes and their neighbors are accessed to get the other place-aspect\_category pairs. For a given aspect category, the neighbor nodes are sorted based on their rank before the explanation is generated.
An explanation of the following form is generated:
\begin{enumerate}
	\item Food Category
	
	Places ordered by rank: Place 1, Place 2, Place 3, ...
	
	\item Service Category
	
	Places ordered by rank: Place 4, Place 5, Place 1, ...
	
	
\end{enumerate} 

\subsection{Dense subgraph extraction}
\label{ssec:methodology_dense_subgraph:num3}
\begin{figure*}[h!]
	\centering
	\subfloat[Overview of user-aspect\_category-location relation]{\includegraphics[width=0.55\columnwidth,height=0.4\columnwidth]{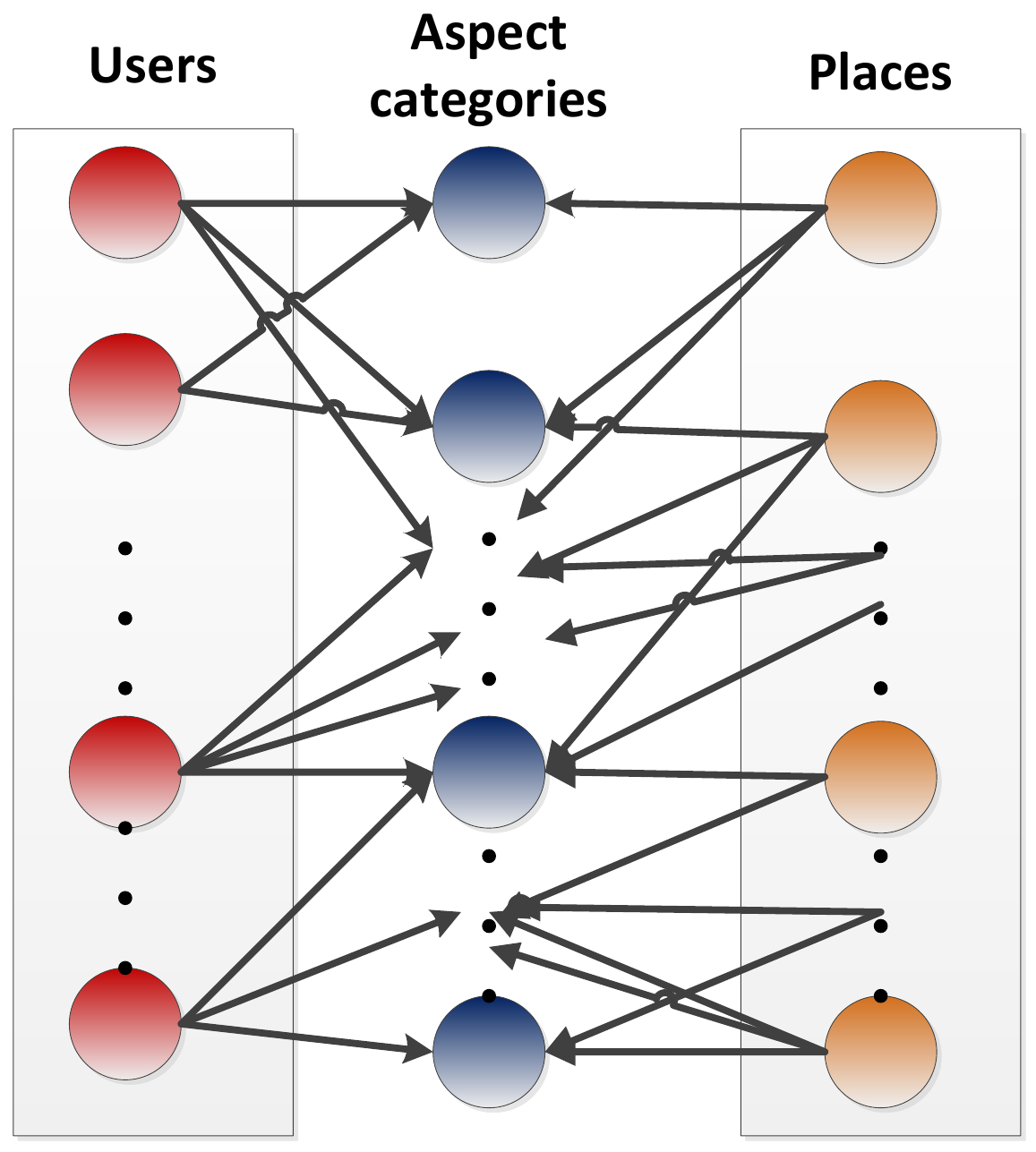}
		\label{fig:bipartite_aspects}}
	\hfil
	\subfloat[Overview of shingles formation]{\includegraphics[width=0.55\columnwidth,height=0.45\columnwidth]{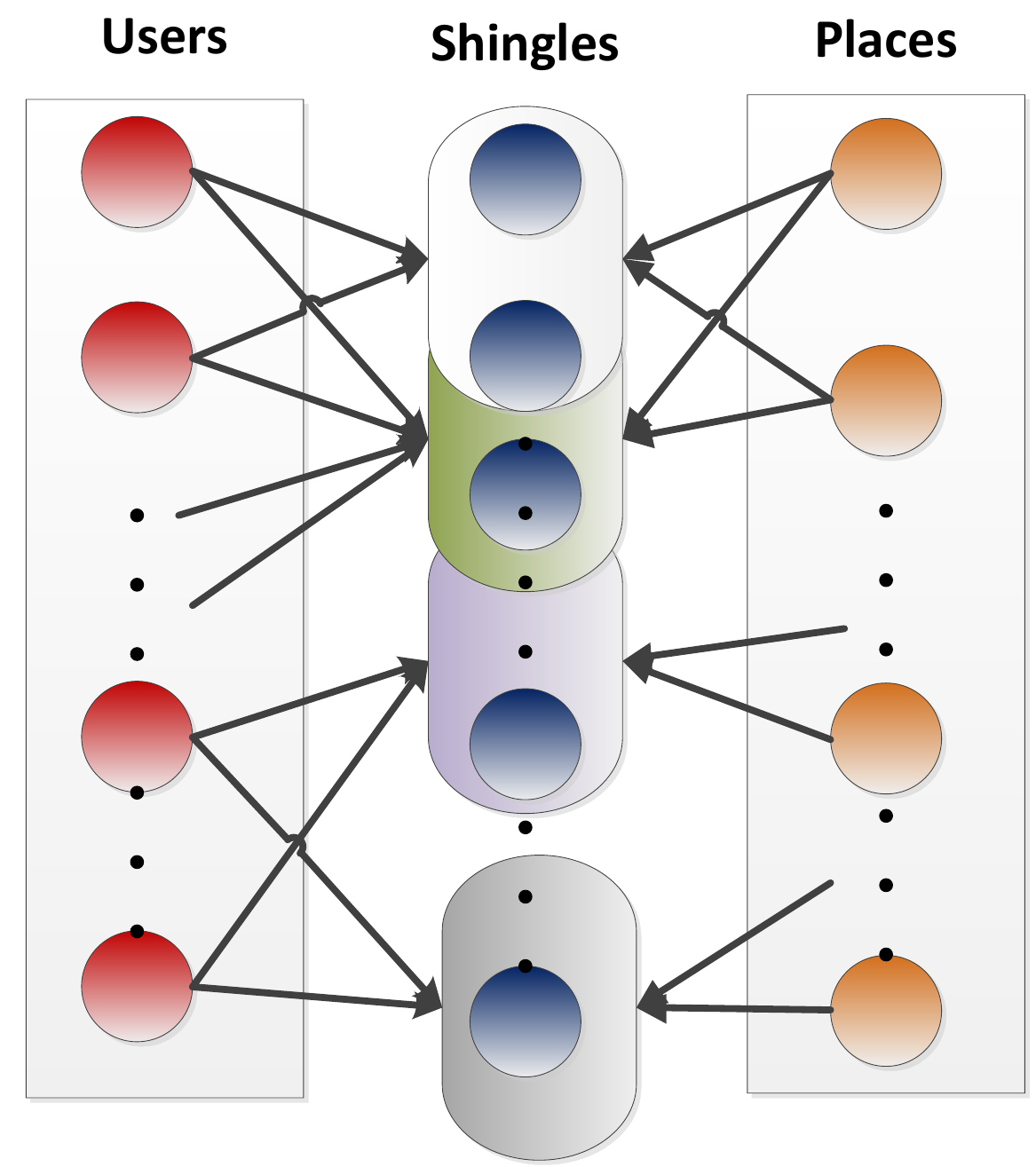}
		\label{fig:bipartite_shingles}}
	\caption{High level overview of dense subgraph extraction (shown without edge weights)}
	\label{fig:shingles_combined}
\end{figure*}

Inspired from~\cite{gibson2005discovering, broder1997syntactic} we represent the user-aspect-location relation as bipartite graphs (user-aspect relation and aspect-location relation). The places recommended to a user, the aspect categories of those places are used to construct a user-aspect, aspect-place bipartite network as shown in Figure~\ref{fig:bipartite_aspects}.
The basic idea is to extract the dense subgraphs from user-aspect bipartite relation and aspect-location bipartite relation. Such dense subgraphs are used to align the user-location tuples that have dense mutual relation through the aspect nodes.
Figure~\ref{fig:shingles_combined} shows a basic representation of the network and the extraction of dense subgraphs.
Unlike a basic bipartite graph, we also exploit the weight of the user-aspect and place-aspect relation because the weights can represent the extent of preference of a user on aspects and the popularity information of a place through the aspects. For instance, a place might be best known for "Food" and might have been rated highly or reviewed more on this aspect but it might be less popular for "Price" aspect and only few visitors might have rated or reviewed on this aspect. It is also highly probable that the reviews of "Food" lover mostly focus on the "Food" aspect category. 
The visitors most likely review on "Food" for the places that is known for "Food" and less likely review on less popular aspect categories. In such a case, having equal weight of "Price" and "Food" aspect might not be the appropriate representation of the place.

We exploit the random extraction of connected components from the network and proceed with the components having high similarity score. If $\gamma$ is a random permutation applied on the homogeneous sets A and B (for instance, set A can contain all user nodes and set B can contain all the aspect category nodes), then the similarity score of the two subsets can be defined as:
\begin{equation}
\text{Sim}^{\gamma}(A,B) = \frac{f(A,B)}{f(A) + f(B)}
\end{equation},
where $f(A,B) = \sum\limits_{\mathclap{\substack{a\in A, b\in B \\ (a,b)\in E}}}W_{a,b}$ gives the contribution of all the edges (a,b), $W_{a,b}$ is the weight of edge (a,b) that is normalized to all the edges outgoing from node a, $f(A) = \sum\limits_{\mathclap{(a,i) \in E}}W_{a,i}$ is the sum of normalized weights of all edges outgoing from node a, and $f(B) = \sum\limits_{\mathclap{(i, b) \in E}}W_{i,b}$ is the sum of normalized weights of all edges incidence on node b. 
This can be generalized to set A and set B of any size. It is obvious that absence of location-aspect edge indicate that the place is not known for that aspect (either the aspect is not relevant or is negatively rated).
We can use the min-wise independent permutations~\cite{broder1997syntactic, broder2000min} to avoid exploitation on each and every permutation to find the sets with high similarity score. As we use some predefined number of permutations, we do not focus on min-wise independence of permutations.

\begin{algorithm}[h!]
	\caption{ShingleFinder(G = (V,E), c, s, k)}
	\label{alg:shingle_finder}
	\begin{algorithmic}[1]
		\STATE {//G is the input graph, V is the set of vertices, and E is the set of edges, c is the number of permutations, s is the length of each set, k is the number of shingles to be extracted}
		\STATE {initialize L as an empty list}
		\FOR{each place node} 
			\FOR{j = 1 \TO c}
				\STATE {get a set of s aspect\_category nodes} 
				\STATE {find aggregated similarity for the place and aspect\_category nodes in this set}
				\STATE {store this set and its score in L}
			\ENDFOR
		\ENDFOR
		\STATE {return k sets with high similarity score (\textit{these sets are called shingles}) from L} 
	\end{algorithmic}
\end{algorithm}

Algorithm~\ref{alg:shingle_finder} defines the steps for extraction of shingles from a bipartite graph. For each place node, we apply ShingleFinder(G, c, s, k) to find the set of aspect\_category nodes linked to it and to extract the k shingles for each place node. For each shingle created, we find the list of all place nodes that contain it. These are the places that are mostly reviewed for the aspect\_category nodes contained in the shingle (see Figure~\ref{fig:bipartite_shingles} for detail). As shingles can contain overlapping set of aspect\_categories, it can represent the places and user preferences of overlapping aspect\_categories as well.
A similar approach is taken to find the shingles for user nodes. The shingles of a user node represent the set of aspect categories that adhere to user preference (the preference can be ordered based on the similarity score of a user node to the shingles).
As our goal is to cluster user, location tuples, we need to find the sets of user and location nodes that share sufficiently large number of shingles. Each shingle contains the associated aspects which can be the means to link the users and locations. We can easily find the top $n_u$ users and top $n_l$ locations whose similarity score is high for this shingle. The place and the aspect\_categories from the most similar shingles are used to generate the explanations. The overall process can be achieved in polynomial time and is dependent on the number of nodes in the graph, number of shingles to use, and the size of a shingle. This model is termed as \textbf{PERS-Dense} in the rest of this paper.

Finding the subsets of aspect\_categories with highest similarity score not only facilitates explanation of recommendation but also provisions clustering of the users who have similar preferences on aspects (even in case of absence of explicit social links) and generate a group recommendation. It can be used to generate preference wise recommendation (e.g., for the set of users \{$u_1, u_2, u_5$\} the set of aspect\_categories \{"food", "service"\} might be interesting, for the set of users \{$u_1, u_2, u_3$\} the set of aspect\_categories \{"food", "price"\} might be interesting, etc.).
This can also facilitate the clustering of places that are preferred for similar aspects (for instance, the cluster of hotels that are popular for some set of aspects).

\section{Evaluation}
\label{sec:evaluation}
We defined four models: (i) \textbf{DAP} - the model that used a deep network for recommendations and has no provision for explanation, (ii) \textbf{PERS-Core} - the model that used bipartite core for explanation generation, (iii) \textbf{PERS-Rank} - the model that used a ranking approach for explanation generation, and (iv) \textbf{PERS-Dense} that exploited a dense subgraph extraction for explanation generation.
We evaluated the individual components (Aspect extraction, Aspect categorization, Sentence-aspect-category classification) in terms of accuracy.
For aspect extraction module and aspect categorization module, we used SemEval 2014 Task 4: Aspect Based Sentiment Analysis
Annotation dataset~\footnote{\tiny{http://alt.qcri.org/semeval2014/task4/index.php?id=data-and-tools}} as the benchmark dataset. Using this dataset as a benchmark seems reasonable because it is quite popular for research relevant to aspect, aspect sentiment, and aspect category extraction from customer reviews in different domains. 
This dataset contains sentences from customer reviews for restaurants and laptops. The review sentences contain annotation for the aspect terms, aspect term polarity, aspect category (e.g., food, service, price, ambience (relevant to the atmosphere and environment of a place), and anecdotes (e.g., the sentences that do not fit in above four categories) ), and aspect category polarity.
\begin{enumerate}
	\item Aspect extraction: We used the SemEval 2014 Task 4: Aspect Based Sentiment Analysis
	Annotation dataset as the benchmark data and were able to get an accuracy of 70\%. The performance was 82\% on all of our evaluation datasets.
	
	\item Aspect categorization: We got an accuracy of 67\% with the SemEval 2014 Task 4: Aspect Based Sentiment Analysis
	Annotation dataset. The performance on aspect categorization on our evaluation dataset was 75\% for Yelp dataset, 77\% for TripAdvisor dataset, and 75\% for Airbnb dataset.
	
	\item Sentence-aspect-category classification: We used 100, 150, and 200 epochs with 32 and 64 batches. With 200 epochs and 64 batches, we got 69\% accuracy on Yelp dataset and 65\% on TripAdvisor and Airbnb dataset. The model labeled $\sim$15\% of sentences as "NONE".
	
\end{enumerate}
We also evaluated the performance of the recommendation generated by our proposed model. 
We compared the performance of our proposed models with the following models:
\begin{enumerate}
	\item Word-embedding approach - In this approach, the review sentences from a user and the one for an item are mapped to a latent space using the word embedding~\cite{mikolov2013distributedword2vec}. For a user, the K-nearest neighbors in the space were considered as the top-K recommendations.
	
	\item Latent Dirichlet Allocation approach~\cite{blei2003latent} - In this model, we extract the topics relevant to a user and the topics relevant to places. The user-place tuples with most common topics are used for the recommendation. 
	
	\item DeepConn~\cite{zheng2017joint} - This CNN-based model uses the review embeddings but ignores other features' embedding and the polarity of reviews.
	
\end{enumerate}

\subsection{Dataset}
\label{subsec:evaluation:dataset}
We use three real-world datasets (see Table~\ref{tab:dataset}) to evaluate the proposed models. Table~\ref{tab:dataset} shows that in all three datasets, most of the users tend to give high (positive) ratings to the places.
\begin{table}[h!]
	\begin{tabular}{|c|c|c|c|}
		\hline 
		\rule[-1ex]{0pt}{2.5ex} \textbf{Attributes} & \textbf{Yelp}\protect\footnotemark & \textbf{TripAdvisor}\protect\footnotemark & \textbf{AirBnB}\protect\footnotemark \\ 
		
		\hline 
		\rule[-1ex]{0pt}{2.5ex}Reviews & 2,225,213 & 246,399 & 570,654\\ 
		
		\hline 
		\rule[-1ex]{0pt}{2.5ex}Users & 552,339 & 148,480 & 472,701\\ 
		
		\hline 
		\rule[-1ex]{0pt}{2.5ex} Places & 77,079 & 1,850 & 26,734\\ 
		\hline 
		\rule[-1ex]{0pt}{2.5ex}Words & 302,979,760 & 43,273,874 & 54,878,077\\ 
		\hline 
		
		\rule[-1ex]{0pt}{2.5ex}Sentences & 18,972,604 & 2,167,783 & 284,1004 \\ 
		\hline
		
		\rule[-1ex]{0pt}{2.5ex} \begin{tabular}{@{}c@{}}
			Avgerage \\ Sentences/review \end{tabular}   & 8.53 & 8.79 & 4.98 \\ 
		\hline 
		
		\rule[-1ex]{0pt}{2.5ex} \begin{tabular}{@{}c@{}}
			Avgerage \\ Words/review \end{tabular}  & 136.15 & 175.62 & 96.16\\ 
		\hline 
		
		\rule[-1ex]{0pt}{2.5ex} \begin{tabular}{@{}c@{}}
			Avgerage \\ Reviews/user \end{tabular}  & 4.03 & 1.66 & 1.20\\ 
		\hline
		
		\rule[-1ex]{0pt}{2.5ex} \begin{tabular}{@{}c@{}}
			Avgerage \\ Reviews/place \end{tabular}  & 28.87 & 133.18 & 21.34\\ 
		\hline
		
		\rule[-1ex]{0pt}{2.5ex}4, 5 stars\protect\footnotemark & \begin{tabular}{@{}c@{}}
			591,618 \\ and 900,940 \end{tabular}   & \begin{tabular}{@{}c@{}}
			78,404 \\  and 104,442 \end{tabular}   & 479,842\\ 
		\hline
		
		\rule[-1ex]{0pt}{2.5ex}1, 2 stars & \begin{tabular}{@{}c@{}}
			260,492 \\  and 190,048 \end{tabular}  & \begin{tabular}{@{}c@{}}
			15,152 \\   and 20,040 \end{tabular}  & 5,766\\ 
		\hline
		
	\end{tabular} 
	\caption{Statistics of the datasets}
	\label{tab:dataset}
\end{table}
\addtocounter{footnote}{-3}
\footnotetext{\tiny{https://www.yelp.com/dataset\_challenge}}
\addtocounter{footnote}{1}
\footnotetext{\tiny{Wang et al.~\cite{wang2011latent}}}
\addtocounter{footnote}{1}
\footnotetext{\tiny{http://insideairbnb.com/get-the-data.html}}
\addtocounter{footnote}{1}
\footnotetext{\tiny{explicitly missing ratings, neutral, and zero ratings are not shown}}
The top-10 terms of different aspect categories are illustrated in Table~\ref{tab:aspect_categories_top}.
\begin{table}[h!]
	\begin{tabular}{|c|c|}
		\hline 
		\rule[-1ex]{0pt}{2.5ex} \textbf{Category} & \textbf{Terms}\\ 
		
		\hline 
		
		\rule[-1ex]{0pt}{2.5ex}  Price & \begin{tabular}{@{}c@{}}
			cash, redeem, cheap, expensive, afford, refund, \\skyrocket, economize, reimburse, discount\end{tabular} \\ 
		
		\hline 
		
		\rule[-1ex]{0pt}{2.5ex}  Food & \begin{tabular}{@{}c@{}} cappuccino, buffet, mushroom, cranberry, salami,\\ healthy, shell, croissant, sushi, broccoli \end{tabular} \\ 
		
		\hline 
		
		\rule[-1ex]{0pt}{2.5ex}  Pet  & \begin{tabular}{@{}c@{}} mew, swan, cat, fish, ant, pony, dog, bird, duck, purr  \end{tabular}   \\ 
		
		\hline 
		
		\rule[-1ex]{0pt}{2.5ex}  Service & \begin{tabular}{@{}c@{}} friendly, repair, employment, discount, servings,\\ safari, checkouts, cleansing, sightseeing, attitude \end{tabular} \\ 
		
		\hline 
		
		
		
		\rule[-1ex]{0pt}{2.5ex}  Amenities & \begin{tabular}{@{}c@{}} breakfast, massage, yoga, housekeeping, excursion, \\gamble, sightseeing, paraglide, exercise, television \end{tabular}  \\ 
		
		\hline 
		
		
		
	\end{tabular} 
	\caption{Top-10 terms in different aspect categories}
	\label{tab:aspect_categories_top}
\end{table}

\subsection{Experimental settings}
\label{subsec:evaluation:exp_settings}
We used 5-fold cross validation to measure the performance of the models. The frequency thresholds for noun and noun phrase extraction were set to 100, 250, and 500. From our experimental analysis, the threshold of 100 was found to be better. For CNN, we used 128 filters, 64 batches, 200 epochs, and embedding vectors of size 384. An Ubuntu 14.04.5 LTS, 32 GB RAM, a Quadcore Intel(R) Core(TM) i7-3820 CPU @ 3.60 GHz was used to evaluate the models. The same configuration with a Tesla K20c 6 GB GPU was used to evaluate the neural network-based models. 

\subsection{Impact of Explanation- A Case Study}
\label{ssec:}
In this subsection, we present an example of the bipartite core detection and the explanation generation. 
We take the users from Yelp dataset and analyze the impact of categorizing the recommended places based on the bipartite cores. We can see how the bipartite core-based model outperforms the simple top@N-based models.

For simplicity, we analyze the impact of explanation on 200 different users and five different aspect categories (see Table~\ref{tab:aspect_categories}). We present illustrations of three users - "7iigQ2XM-V0ciwmCIdrIBA", "7Mg6r6g7RUwQH\_Bllrd-wQ", and "9HDElil2309UajBgtYcD4w", hereafter known as $u_1$, $u_2$, and $u_3$ respectively. We take the recommendations from \textbf{DAP} model as the simple recommendation for these users. A bipartite graph was created with the places recommended (which can be taken from \textbf{DAP} or any other models) to these users and the relevant aspect categories (which were positively reviewed (in majority)) of those places. A directed edge was created between aspect\_category and place if the review of the place was labeled for this aspect category.

Table~\ref{tab:bipartitesummary} illustrates the summary of top-5 bipartite cores extracted for these users. 
\begin{table}[h!]
	\begin{tabular}{|c|c|c|c|}
		\hline 
		\rule[-1ex]{0pt}{2.5ex} \textbf{Bipartite Cores} & \textbf{User} $u_1$ & \textbf{User} $u_2$ & \textbf{User} $u_3$ \\ 
		
		\hline 
		\rule[-1ex]{0pt}{2.5ex}First core & 
		
		\begin{tabular}{@{}c@{}}
			\textit{Price}
			\\ \hline
			\\103 places 
		\end{tabular}
		& 
		\begin{tabular}{@{}c@{}}
			\textit{Service}
			\\ \hline
			\\137 places 
		\end{tabular}
		& 
		\begin{tabular}{@{}c@{}}
			\textit{Price}
			\\ \hline
			\\272 places 
		\end{tabular}
		\\ 
		
		\hline 
		\rule[-1ex]{0pt}{2.5ex}Second core & 
		\begin{tabular}{@{}c@{}}
			\textit{Pet}
			\\ \hline
			\\103 places 
		\end{tabular}
		& 
		\begin{tabular}{@{}c@{}}
			\textit{Price}
			\\ \hline
			\\137 places 
		\end{tabular} 
		& 
		\begin{tabular}{@{}c@{}}
			\textit{Service}
			\\ \hline
			\\272 places 
		\end{tabular}\\ 
		
		\hline 
		\rule[-1ex]{0pt}{2.5ex} Third core 
		& 
		\begin{tabular}{@{}c@{}}
			\textit{Service}
			\\ \hline
			\\47 places 
		\end{tabular}
		& 
		\begin{tabular}{@{}c@{}}
			\textit{Pet}
			\\ \hline
			\\137 places 
		\end{tabular} 
		& 
		\begin{tabular}{@{}c@{}}
			\textit{Pet}
			\\ \hline
			\\272 places 
		\end{tabular}\\ 
		
		\hline 
		\rule[-1ex]{0pt}{2.5ex}Fourth core 
		& 
		\begin{tabular}{@{}c@{}}
			\textit{Food}
			\\ \hline
			\\103 places 
		\end{tabular}
		& 
		\begin{tabular}{@{}c@{}}
			\textit{Food}
			\\ \hline
			\\42 places 
		\end{tabular} 
		& 
		\begin{tabular}{@{}c@{}}
			\textit{Food}
			\\ \hline
			\\1 place 
		\end{tabular}\\ 
		\hline 
		
		\rule[-1ex]{0pt}{2.5ex}Fifth core 
		& 
		\begin{tabular}{@{}c@{}}
			\textit{Amenities}
			\\ \hline
			\\9 places 
		\end{tabular}
		& 
		\begin{tabular}{@{}c@{}}
			\textit{Amenities}
			\\ \hline
			\\137 places 
		\end{tabular}
		& 
		\begin{tabular}{@{}c@{}}
			\textit{Amenities}
			\\ \hline
			\\81 places 
		\end{tabular}
		\\ 
		\hline
		
	\end{tabular} 
	\caption{Summary of bipartite cores for different users}
	\label{tab:bipartitesummary}
\end{table}
From Table~\ref{tab:bipartitesummary}, we can see that the ordered preferences of user $u_1$ are {"Price", "Pet", "Service", "Food", and "Amenities"}. On the other hand user $u_2$ has {"Service", "Price", "Pet", "Food", and "Amenities"}. This shows that when selecting places, regardless of the order of places the simple recommendation provided, the user $u_1$ prefers the places that were popular for "Price" category. 
The authority score of nodes can be used to order places within a given aspect category.
The findings of our analysis are summarized below:
\begin{enumerate}
	\item For the user $u_1$, the place "NK3S3U6TQtysH\_-eqT3bBQ" was the second highly recommended place by simple recommendation. With the bipartite core, it is categorized into "Others" bipartite core which in-fact is the sixth core. So, if the user really cares about other cores (i.e. related to other aspect categories) then having it in sixth core is better than having it in front list. 
	
	The location "p9Bl3BxPltz2WnIxJLnBvw" which was the least recommended one with simple recommendation (without bipartite core based explanation) is now categorized as the least popular item for the primary bipartite core (i.e. related to "Price"), and three other secondary cores (i.e. related to "Service", "Pet", and "Food"). Many places ranked in the later part of the list by the simple recommendation were found within top-20 of the different bipartite cores. Have this user used the simple recommendation and considered only the top-20 recommendations then these items would have been missed.
	
	\item The place "v4iA8kusUrB19y2QNOiUbw" which was the top recommended item for user $u_2$ by the simple recommendation is categorized to the sixth bipartite core (i.e. "Others"). The place "HxPpZSY6Q1eARuiahhra6A" that did not fit in top-20 of simple recommendation is found in the sixth position of first three bipartite cores. 
	
	The location "mh1le9QGMrZLohAjfheJJg" which was the second least recommended by simple recommendation is categorized as the second least preferred item for the first five bipartite core (i.e. "Service", "Price", "Pet", "Food", and "Amenities").
	
	\item For user $u_3$, the place "d4JotJ\_TGNCm9y-TB4e86Q" which was top second recommended place is categorized into sixth core (i.e. "Others"). On the other hand, the place "y9cQ0DBC0qFNgkpTXHzokA" which was least recommended by the simple recommendation is found in the first four bipartite cores (i.e. "Price", "Service", "Pet", and "Food") relevant to her. 
	
	The place "Jp8Tz0\_OK3T71eYPmC2qww" which was ranked beyond the top-20 by simple recommendation is ranked within the top-10 locations for the primary bipartite core.
\end{enumerate}

This implies that simple recommendation score is not always reliable when the preferences are categorized into different aspects. As the items at front of recommended list might be least preferred for some aspect, simply taking top@N items might adversely impact the performance of recommendation.
Using our approach, a sample explanation for a user is the ordered set of places taken from the ordered bipartite cores. For e.g., for $u_1$ it can be: 
\begin{enumerate}
	\item Place recommended: Place 1, Place 2,...
	
	Explanation: Popular for best price.
	
	\item Place recommended: Place 3, Place 4,...
	
	Explanation: Popular due to pet friendliness.
\end{enumerate}

\subsection{Evaluation of Explainability}
We checked the presence of correct aspects in the explanation. For this, we ordered the aspects of every place based on the aspect popularity score for that place. For a place \textit{p}, the aspect popularity of an aspect \textit{a} can be defined in terms of the number of positive and negative mentions of that aspect and is defined as:
\begin{equation}
Aspect\space Popularity(p_a) = \sum\limits_{\mathclap{sentence \in Review_{p}}} (\mid positive \mid a - \mid negative \mid a).
\end{equation}
We used a trigram across the extracted aspects to identify the sentiment polarity of the aspects.
For every place, the relevant aspects were ordered by the aspect popularity score. So, a place can be represented by the set of aspects ordered by their popularity score: 
$p_a =$ \textbraceleft $a_1, a_2,....,a_n $\textbraceright. 
For every explanation, we took the aspects for which a place was recommended. The aspects were ordered based on the order of cores (primary, secondary, etc.). This gave us another set of aspects for every place.
The performance of explainability was then measured in terms of Levenshtein distance (i.e. edit distance) between the lists. The average Levenshtein distance across all the places was observed to be 20\% with PERS-Core. Similarly, the highly ranked items in PERS-Rank and the denser subgraphs from PERS-Dense were used to extract the set of places  and their aspects. Such set of aspects were compared to the set of aspects obtained by ordering their frequency of mentions. The deviation of such two lists were observed as 25\% for PERS-Rank and 27\% for PERS-Dense. This shows that with small deviation, the aspects-based explanation maintains the preferences explicitly shared via user reviews.

\subsection{Experimental Results and Discussion}
\label{subsec:evaluation:results}
\begin{figure*}[!t]
	\centering
	\subfloat[Precision@N on Yelp Dataset]{\includegraphics[width=0.65\columnwidth,height=0.45\columnwidth]{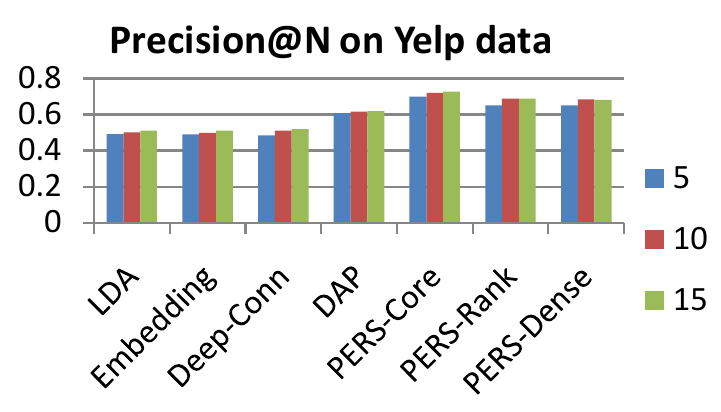}
		\label{fig_first_case}}
	\subfloat[Precision@N on TripAdvisor Dataset]{\includegraphics[width=0.65\columnwidth,height=0.45\columnwidth]{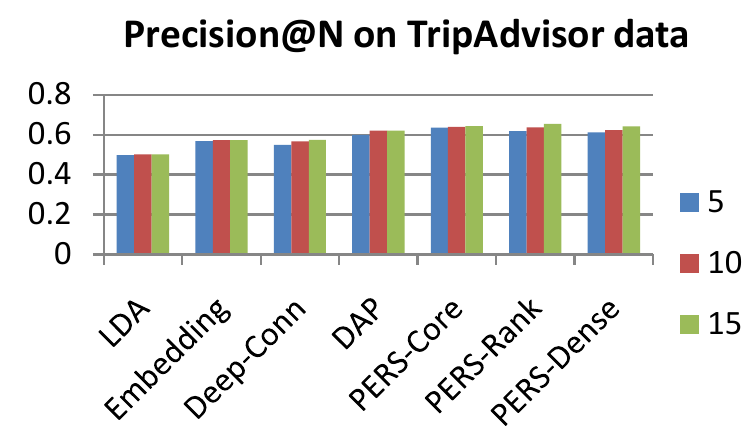}
		\label{fig_second_case}}
	\subfloat[Precision@N on Airbnb Dataset]{\includegraphics[width=0.65\columnwidth,height=0.45\columnwidth]{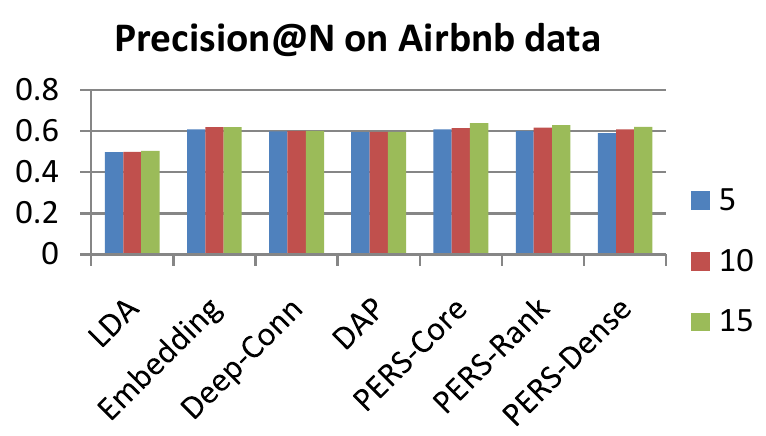}
		\label{fig_second_case}}
	\caption{Precision of different models}
	\label{fig:res_prec}
\end{figure*}

\begin{figure*}[!t]
	\centering
	\subfloat[Recall@N on Yelp Dataset]{\includegraphics[width=0.65\columnwidth,height=0.45\columnwidth]{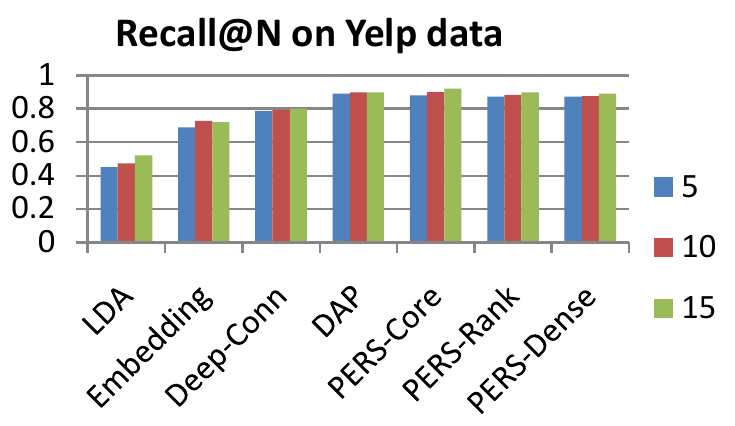}
		\label{fig_first_case}}
	\subfloat[Recall@N on TripAdvisor Dataset]{\includegraphics[width=0.65\columnwidth,height=0.45\columnwidth]{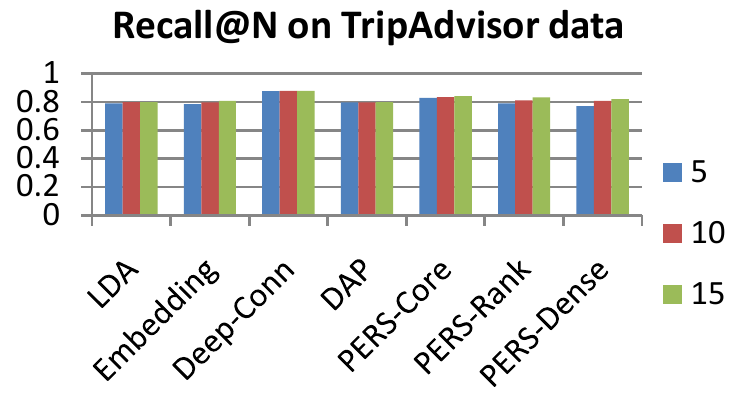}
		\label{fig_second_case}}
	\subfloat[Recall@N on Airbnb Dataset]{\includegraphics[width=0.65\columnwidth,height=0.45\columnwidth]{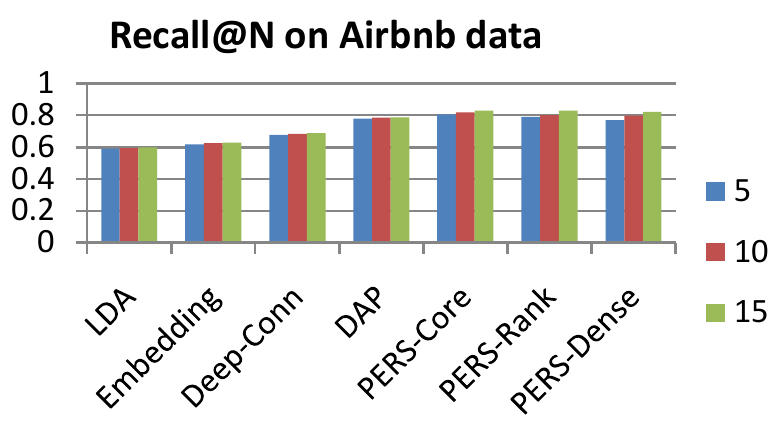}
		\label{fig_second_case}}
	\caption{Recall of different models}
	\label{fig:res_rec}
\end{figure*}

The chronologically sorted reviews of users and places with at least five reviews were used in the evaluation.
We used a 5-fold cross validation and the precision (p), recall (r), and f-score (2*p*r/(p+r)) metrics to evaluate the models. We considered the cases of top @5, @10, @15, and @20 recommended items for the evaluation. The evaluation of different models on Yelp dataset, TripAdvisor dataset, and Airbnb dataset is illustrated in~Table~\ref{tab:results_yelp}, \ref{tab:results_trip}, and \ref{tab:results_air} respectively. The performance of different models in the form of Precision@N and Recall@N is illustrated in Figure~\ref{fig:res_prec} and Figure~\ref{fig:res_rec}.

\begin{table}[h!]
	\centering
	\begin{tabular}{|c|c|c|c|}
		\hline \textbf{Models} & \textbf{Precision}& \textbf{Recall} & \textbf{F-Score}\\
		\hline \textbf{Yelp Dataset}\\
		
		
		\hline \textit{LDA~\cite{blei2003latent}} & 0.5016 & 0.4828 & 0.4920  \\
		
		
		\hline \textit{Embedding~\cite{manotumruksa2016modelling}} & 0.5002 & 0.7125 & 0.5878  \\
		
		\hline \textit{DeepConn~\cite{zheng2017joint}}  & 0.5051 & 0.7935 & 0.6172  \\
		
		\hline \textit{DAP} & 0.6155 & 0.8963 & 0.7298  \\
		
		\hline \textbf{\textit{PERS-Core}} & \textbf{0.7168} & \textbf{0.8996} & \textbf{0.7978$^*$}  \\
		
		\hline \textit{PERS-Rank} & 0.6774 & 0.8842 & 0.7671  \\
		
		\hline \textit{PERS-Dense} & 0.6731 & 0.8794 & 0.7625  \\
		
		
	\hline
	\end{tabular}
	\caption{Average Performance of models on Yelp dataset}
	\label{tab:results_yelp}
	\centering
	\end{table}

\begin{table}[h!]
	\centering
	\begin{tabular}{|c|c|c|c|}
		\hline \textbf{Models} & \textbf{Precision}& \textbf{Recall} & \textbf{F-Score}\\
		\hline \textbf{TripAdvisor Dataset} \\
		
		
		\hline \textit{LDA~\cite{blei2003latent}} & 0.5000 & 0.7968 & 0.6144  \\
		
		
		\hline \textit{Embedding~\cite{manotumruksa2016modelling}} & 0.5711 & 0.7971 & 0.6654  \\
		
		\hline \textit{DeepConn~\cite{zheng2017joint}} & 0.5634 & \textbf{0.8781} & 0.6864  \\
		
		\hline \textit{DAP} & 0.6131 & 0.7988 & 0.6937  \\
		
		\hline \textbf{\textit{PERS-Core}} & \textbf{0.6388} & 0.8341 & \textbf{0.7235$^*$}  \\
		
		\hline \textit{PERS-Rank} & 0.6366  & 0.8112 & 0.7133  \\
		
		\hline \textit{PERS-Dense} & 0.6254 & 0.7998 & 0.7019  \\
		
		
		\hline
	\end{tabular}
	\caption{Average Performance of models on TripAdvisor dataset}
	\label{tab:results_trip}
	\centering
\end{table}

\begin{table}[h!]
	\centering
	\begin{tabular}{|c|c|c|c|}
		\hline \textbf{Models} & \textbf{Precision}& \textbf{Recall} & \textbf{F-Score}\\
		\hline \textbf{AirBnB Dataset}\\
		
		
		\hline \textit{LDA~\cite{blei2003latent}} & 0.5000 & 0.5948 & 0.5433  \\
		
		
		\hline \textit{Embedding~\cite{manotumruksa2016modelling}} & 0.6164 & 0.6243 & 0.6203 \\
		
		\hline \textit{DeepConn~\cite{zheng2017joint}} & 0.6001 & 0.6832 & 0.6389  \\
		
		\hline \textit{DAP} & 0.5972 & 0.7845 & 0.6781  \\
		
		\hline \textbf{\textit{PERS-Core}} & \textbf{0.6216} & \textbf{0.8183} & \textbf{0.7065$^*$}  \\
		
		\hline \textit{PERS-Rank} & 0.6161 & 0.8073 & 0.6988  \\
		
		\hline \textit{PERS-Dense} & 0.6077 & 0.7970 & 0.6896  \\
		
		\hline
	\end{tabular}
	\caption{Average Performance of models on Airbnb dataset}
	\label{tab:results_air}
	\centering
\end{table}


From the evaluation results, we can see that among the ones without explanation, \textbf{DAP} performed best on the Yelp dataset. Though it outperformed other models in other two datasets as well, the difference was not that much significant. This implies that for larger datasets, the performance of the proposed model is outstanding. 
This is common with DNNs which need a reasonable amount of training data for better performance. 
The recall of DeepConn~\cite{zheng2017joint} was higher than that of \textbf{DAP} in the TripAdvisor dataset but its precision and F-Score were lower than that of \textbf{DAP}. This might be because of the sentence-level sentiment which was exploited in DAP but not in DeepConn~\cite{zheng2017joint}.

Unlike \textbf{DAP}, which provided a single list of recommendations and selected top@N locations from the list, the \textbf{PERS-Core} and \textbf{PERS-Rank} produced individual lists for each aspect category.
These two models performed better than \textbf{DAP} because they categorized recommendations into different aspect categories which led to the re-ordering of the items into small recommendation lists. This re-ordering can help increase the number of true positives and decrease the false positives, as the least preferred items might move to the later part of the recommended lists and the more preferred ones move to the front part of the lists. The performance of \textbf{PERS-Core} was slightly better than \textbf{PERS-Rank}. One reason might be due to the repeated bipartite core extraction by \textbf{PERS-Core} where the nodes got re-ranked for every bipartite core but the \textbf{PERS-Rank} only ranked all the nodes just once.

The performance of \textbf{PERS-Dense} was in par with \textbf{PERS-Rank}. The small difference might be due to the categorical context incorporated in \textbf{PERS-Rank} but not in \textbf{PERS-Dense}. As we have five different aspect\_categories, we evaluated three different (s=\{3,4,5\}) length of shingles. 
We used the value of c (the number of combinations) to include all the possible unique combinations of aspect\_categories (e.g., we used c=10 for s=3).
We used k=\{3, 5, 7\} as the number of shingles to cluster and compare users and locations. The best performance was observed when s=3, c=10, and k=5. As considering more shingles (lots of aspect combinations) might not be able to exploit the personalized preference of users sharing all those shingles, performance on k=7 was lower than that of k=5. The best performance on k=5 implies that a reasonable number of aspect combinations can better model the extent of preference of users. 

After having the ordered set of places within each aspect category, an explanation of type similar to~\cite{lawlor2015opinionated} (i.e. place A is better than 80\% of places for "Food", and so on) can be simply achieved by counting the number of places behind the target place in the recommended list. 
The application of additional text (or text templates) to adorn the explanations can be achieved with simple implementation enhancements. 
\subsection{Limitations} The deep neural networks require lots of training data for efficient results. The prediction relies on the data used in off-line training and might need further tunings to handle real-time data. The model relied on single approach to extract the aspects. The model can also be extended to exploit phrase-level aspect and sentiment analysis which can capture the multi-aspect multi-sentiment relation within a sentence.

\section{Conclusion and Future Work}
\label{sec:conclusion}
We formulated the review-aspect\_category correlation using deep neural network and represented the place-aspect\_category bipartite relation as a bipartite graph. We proposed a bipartite-core based, ranking-based, and a shingles-based method to extract the dense subgraphs that represented the users' ordered preferences on aspect categories. The aspect categories from the dense subgraphs were used to explain the recommendation.
The evaluations on three real-world datasets and the presented case studies clearly demonstrate the efficiency of proposed models. 
\textbf{PERS-Dense} can also be used to generate explainable recommendations to a group of users. Besides simple explanation of recommendation, the shingles can be used for additional purposes, such as (i) to generate overlapping cluster of places that are known for a set of aspect\_categories, (ii) generate preference-wise recommendation to a preference-based group, etc.
In future, we would like to cluster the users based on their preference order on aspect categories, incorporate more aspect categories, evaluate the explanations with additional metrics, and exploit the dynamic user-aspect preferences. 

\bibliographystyle{IEEEtran}
\bibliography{explainable_reco_ref}
%
%


\end{document}